\documentclass[]{WileyChemistry-template}

\title{\raggedright Ab-initio force prediction for single molecule force spectroscopy made simple}

\author{
\begin{minipage}{\textwidth}
	Pooja Bhat,\textsuperscript{[a,b]} Wafa Maftuhin,\textsuperscript{[a, c]} Michael Walter,*\textsuperscript{[a,b,d]}
\end{minipage}
}

\newcommand{\affiliation}{
\begin{itemize}

\item[{[a]}] P. Bhat, Dr. W. Maftuhin, Dr. M. Walter\\
FIT Freiburg Centre for Interactive Materials and Bioinspired Technologies,
University of Freiburg, Freiburg, Germany.\\
E-mail: Michael.Walter@ifmf.uni-freiburg.de

\item[{[b]}] P. Bhat, Dr. M. Walter\\
Cluster of Excellence livMatS @ FIT, Freiburg, Germany.

\item[{[c]}] Dr. W. Maftuhin\\
Universitas Negeri Surabaya, Surabaya, Indonesia.

\item[{[d]}] Dr. M. Walter\\
Fraunhofer IWM, MikroTribologie Centrum $\mu$TC,
Freiburg, Germany.

\item[{[\texttt{+}]}] These authors contributed equally.
\end{itemize}
}

\renewcommand{\dedication}{
	\begin{minipage}{\textwidth}
	
	\end{minipage}
}

\renewcommand{\abstract}{
Bond rupture under the action of external forces is induced by temperature fluctuations.
We show that measured forces from single molecule force spectroscopy experiments can be predicted from two quantities describing the
bond that are the barrier to break the bond in absence of force as well as the maximal force the bond can withstand.
The former can be obtained by a force free transition state calculation and
the latter is determined by a simple constrained geometry simulates forces (COGEF)
calculation. 
Considering experimental temperature and force loading rate allows the prediction
of measured bond rupture forces from a closed expression with very good accuracy.
}

\newcommand{\keywords}{
	Mechanochemistry \textbullet\ 
	COGEF \textbullet\ 
	Force Spectroscopy \textbullet\ 
	Bond Breaking 
}

\usepackage{xcolor}

\begin{document}

\twocolumn[\vspace{-1.5cm}\maketitle\vspace{-1cm}
\textit{\dedication}\vspace{0.4cm}]
\small{\begin{shaded}
		\noindent\abstract
	\end{shaded}
}

\begin{figure} [!b]
\begin{minipage}[t]{\columnwidth}{\rule{\columnwidth}{1pt}\footnotesize{\textsf{\affiliation}}}\end{minipage}
\end{figure}




\section*{Introduction}
\label{introduction}

Mechanochemistry is the change or modification of chemical bonds through
the action of mechanical forces.
It allows for modification of materials via the application of
external forces which are usually acting in the form of external stress.
External forces offer the possibility to enable
reactions that are improbable or not possible by other means, such as
thermal or optical activation alone\cite{yang_phosphate-enabled_2025,do_mechanochemistry_2017}.
Examples are the force induced change of ceiling temperature
in a polymer\cite{hsu_mechanochemically_2023} or
enabling reactions not thermally allowed by Woodward-Hoffmann rules\cite{Wang2014, Lenhardt2010, Wang2015}.

Mechanochemistry acts on the molecular level, where bonds are broken or
reformed, but
mechanical forces are often applied to materials in a rather
unspecific methods like ball milling or ultra-sonification,
where the force acting on specific bonds within the material are
hard to control. 
Also the modification of molecular bonds by stretching of polymeric
material\cite{Kempe2018} and the applictaion of rheological methods\cite{walter_mechanochemical_2023}
is complex due to the not well understood force distribution in complex material, despite resent advantages in this direction\cite{hertel_mechanistically_nodate,traeger_microscopic_2023,raisch_determining_2022}.

The cleanest experimental observation of mechanochemical effects is given
by measuring molecular forces in single molecule force spectroscopy (SMFS)
experiments\cite{Grandbois1999,Wang2014,klukovich_backbone_2013,faza_solvolytic_2004,Wang2015,Wang2014a,Wang2015a,cai_angle-dependent_2022}. Here single polymer chains are stretched and force plateaus are observed when bonds in monomers are broken\cite{Wang2015}.
It is well understood that these plateaus reflect properties of the monomers experiencing the bond break\cite{Akbulatov2012}.
The numerical determination of force dependent barriers has been successfully employed by several groups\cite{ong_first_2009, dopieralski_role_2011, dopieralski_FTFES_2011, wollenhaupt_should_2015, wollenhaupt_force-induced_2018}
and it was shown that experimental
force-extension curves can be reproduced based on properties determined from ab-initio calculations\cite{Akbulatov2012,Wang2016}.
Interestingly, often the properties of the monomer are sufficient to 
characterize the measured rupture forces\cite{Akbulatov2012}.

The most straightforward computational strategy 
to describe the effect of forces within calculations
is probably the constrained geometry simulates
forces (COGEF) method developed by Beyer\cite{Beyer2000,Ribas-Arino2012}. 
It is therefore applied in many mechanochemical investigations\cite{hemmer_heterolytic_2021,hemmer_triarylmethane_2023,Klein2020}.
The prediction of measured forces directly from COGEF has not been 
successful\cite{Klein2020}, as the forces obtained are up to an order of magnitude larger than the
measured ones\cite{wick_evaluating_2022}.
The underlying reason is the lack of consideration of temperature\cite{walter_mechanochemical_2023,Grandbois1999,Pobelov2017,Hanke2006} as 
it is the temperature fluctuations that allow to overcome the barriers for
bond breaking\cite{khodayeki_force_2022}.
Due to the stochastic nature of these fluctuations,
only most probable forces for bond rupture can be given.
The determination of most probable forces
has a long history and several expressions
were derived mainly to analyse bond rupture or friction experiments\cite{garg_escape-field_1995,Hummer2003,dudko_intrinsic_2006,bullerjahn_theory_2014,friddle_interpreting_2012,cai_anisotropic_2023}.
Expressions for most-probable forces were given by Garg\cite{garg_escape-field_1995} in a very general form.
These were used then used Hummer, Dudko and co-workers\cite{dudko_intrinsic_2006}
Re-binding effects occuring at low applied forces and slow pulling speed were considered also\cite{friddle_interpreting_2012,bullerjahn_reversible_2022}

Here we address the question how the most probable force measured in experiment can be
predicted from ab-initio calculations. We
show that the experimentally measured forces is mainly determined by 
two quantities describing the bond: 
its dissociation energy and the maximal force the bond can withstand.
These two quantities can be obtained from 
ab-initio calculations of the monomer.

The manuscript is organized as follows. We first formulate the theory and assumptions underlying our description of the bond ruptures processes, where we detail 
the properties of the force dependent probability functions determining the
most probable force. The effects of the speed of force increase
(the loading rate) and temperature are discussed.
We the apply our description to published SMFS data.
The paper ends with conclusions.

\section*{Results and Discussion}
\label{results_discussion}

\subsection*{Theory}

\begin{figure}[!h]
    \centering
\includegraphics[width=0.9\textwidth]{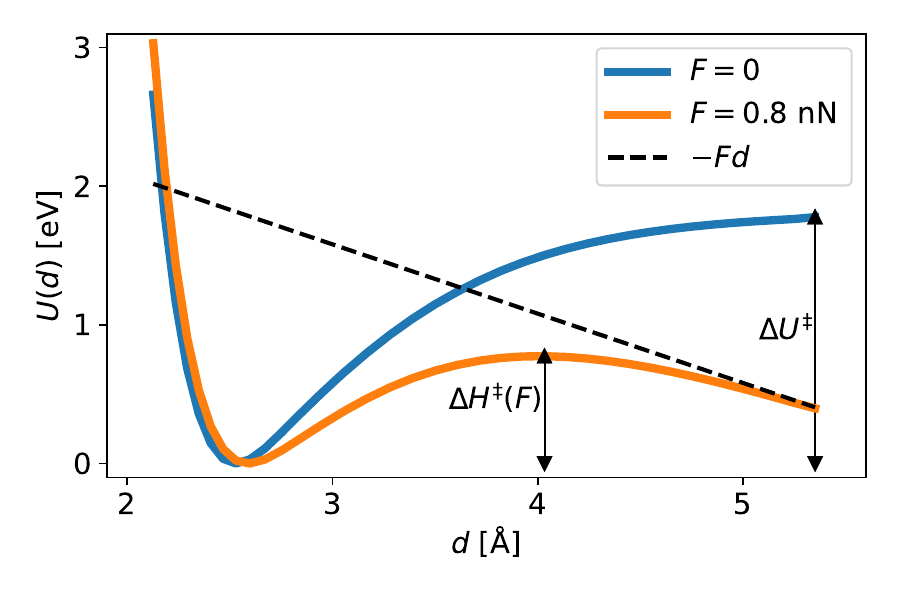} 
    \caption{Force dependent potentials of the AuAg$_2$ model sketched in Fig.~\ref{fig:AuAg2_barrier}. The barriers for bond rupture without force $\Delta U^\ddagger=\Delta H^\ddagger(0)$ 
    and with force $\Delta H^\ddagger(F)$ are indicated. 
    }
    \label{fig:AuAg2_potential}
\end{figure}
We first consider an isolated bond of length $d$
to develop our description. 
We treat bond breaking in a probabilistic picture
in the spirit of Evans and Richie\cite{evans_dynamic_1997} where
the bond potential $U(d)$ is modified by 
the external force as illustrated in Fig.~\ref{fig:AuAg2_potential}. 
The force $F$ is aligned to the direction of the bond (the bond can rotate freely)\cite{Stauch2016}.
Subtracting the work performed on the molecule by the force $F$ 
leads to the enthalpy\cite{khodayeki_force_2022} 
\begin{equation}
    H(F)=U(d) - F d
    \label{eq:H_F}
\end{equation} 
from which the force 
dependent barrier $\Delta H^\ddagger(F)$ (i.e. the force dependent energy difference between the bound initial state and the transitions state) can be obtained.
The restriction to a given external force is a change in variables, such that
the equilibrium bond length $d_0$ is not a free variable anymore, 
but gets a function of $F$, 
i.e. $d_0=d_0(F)$. 
This is reflected by the slight shift of the minimum for finite $F$ as compared
to $F=0$ in the potentials shown in Fig.~\ref{fig:AuAg2_potential}.
The force free energy energy needed to open the bond $\Delta U^\ddagger$
transfers to a barrier $\Delta H^\ddagger(F)$ for finite force.

Apart from the bond-strength $\Delta U^\ddagger$ often considered, 
there is another fundamental property of the bond  $F_{\mathrm{max}}$.
This is the maximal force it can withstand, i.e. the maximal spatial derivative
of $U(d)$. 
It turns out, that the force dependence of the barrier 
$\Delta H^\ddagger(F)$ for many potentials 
can be expressed to good approximation 
in the form\cite{Hanke2006,khodayeki_force_2022}
\begin{equation}
    \Delta H^\ddagger(f) = \Delta U^\ddagger \left( 1 - f\right)^2 ,
    \label{eq:quadratic}
\end{equation}
where $f=F/F_{\mathrm{max}}$.
This is the barrier that results from a cusp-like 
cut quadratic potential considered by many authors\cite{Hummer2003,dudko_intrinsic_2006,bullerjahn_theory_2014,friddle_interpreting_2012}. 
Linearizing Eq.~(\ref{eq:quadratic}) to the small force limit leads 
to the famous Bell barrier\cite{Bell1978,Ribas-Arino2012}
\begin{equation}
    \Delta H^\ddag(F) = \Delta U^\ddag - F x^\ddag
    \label{eq:Bell}
\end{equation}
where we identify the Bell length as $x^\ddag=2\Delta U^\ddagger / F_{\mathrm{max}}$.
This length constant is often interpreted as the distance 
between initial and transition state\cite{avdoshenko_finding_2015}, despite that the latter depends on the force $F$ itself\cite{khodayeki_force_2022}.
Here, $x\ddag$ is purely defined by the two constants of the potential $U^\ddagger$ and $F_{\mathrm{max}}$.

\begin{figure}[!h]
    \centering
\includegraphics[width=\textwidth]{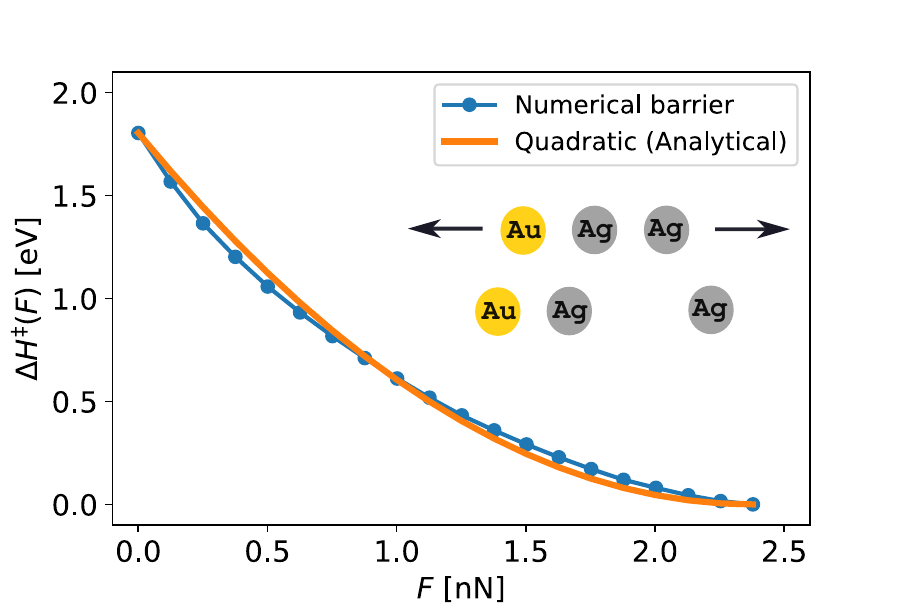} 
    \caption{Barriers for breaking of the Ag-Ag bond under the 
    constraint of a given external force $F$.
    The numerical barrier (determined by an explicit inclusion
    of the external force) is compared to the analytical expression of Eq.~(\ref{eq:quadratic}).
    The setup of the AuAg$_2$ model is sketched.
    }
    \label{fig:AuAg2_barrier}
\end{figure}
We now use the approximate form of the barrier in Eq.~(\ref{eq:quadratic})
to derive the most probable force for bond rupture measured in the experiment under the assumption of a constant loading rate (rate of force increase).
In order to illustrate the steps taken,
we consider a simple linear AuAg$_2$ toy molecule (depicted in Fig.~\ref{fig:AuAg2_barrier})
described by its potential energy obtained from effective medium theory
as implemented in the atomic simulation environment (ASE)\cite{larsen_atomic_2017}. 
The potential in Fig.~\ref{fig:AuAg2_potential} is that of
this model molecule where $d$ is the length of the weaker Ag-Ag bond
and the force $F$ is applied on the two silver atoms directly.

It is practically impossible to apply an external force directly on
two bound atoms in reality. There are always other atoms and bonds
mediating the external force to the bond that ruptures.
We mimic this fact in a very simplistic picture 
by the additional Au atom representing the environment through which
the force is transferred.
When a force acts 
on the outer atoms, the atoms respond by arranging into
a new minimum energy configuration.
This situation can be simulated by displacing 
the outer atoms using the COGEF strategy and relaxing all other degrees of freedom.
This in turn gives the corresponding force needed to stabilize the given outer length
$d$. 
Increasing the outer distance 
eventually leads to rupture of the Ag-Ag bond 
for $\Delta U^\ddagger=$1.80 eV in our example.
The maximal outer force appearing in this process is 
$F_{\mathrm{max}}=$2.38 nN.
These values completely define the form of the barrier 
according to Eq.~(\ref{eq:quadratic}).

Alternatively, we may also determine the force dependent
barrier using the external force is explicitly included (EFEI)
strategy\cite{Ribas-Arino2009,Stauch2016}, where the potential at a fixed external force as in Fig.~\ref{fig:AuAg2_potential} is analyzed.
This strategy leads to a barrier that interpolates between
$\Delta H^\ddagger(F=0) = \Delta U^\ddagger$ and $\Delta H^\ddagger(F_{\mathrm{max}})=0$.
Its form is closely resembled by the quadratic barrier from
Eq.~(\ref{eq:quadratic}).

The analytic expression from Eq.~(\ref{eq:quadratic})
may therefore be used to derive additional analytic results to predict bond rupture at nonzero temperature.
Bond breaking induced by thermal fluctuations is
a probabilistic process.
Denoting the probability that the bond is intact as $P$
and 
assuming first order transitions\cite{dudko_intrinsic_2006}, $P$ follows
\begin{equation}
  \frac{d P}{d t} = -k P  + k_r [1-P ]\; ,
  \label{eq:dPdt}
\end{equation}
where $k$ is the transition rate for 
bond breaking (depending on
the barrier discussed above) and $k_r$ is the rate for bond reforming\cite{friddle_interpreting_2012}. 
All quantities depend on time $t$ either directly or indirectly which is suppressed for clarity here.
We restrict to strong bonds with $\Delta U^\ddagger\gg k_BT$ as is typical for chemically bonded species addressed in SMFS experiments.
This means that the bond is intact ($P=1$) at $t=0$ when no force is applied.
The term involving $k_r$ can be neglected for $1-P\ll P$ (the condition for initial times and strong bonds) or when $k_r \ll k$ as expected for large forces occurring at later times. 
We will therefore set $k_r=0$ in what follows.
A similar approach is taken by Garg\cite{garg_escape-field_1995}, where $P(t)$ is denoted as $W(t)$, and Hummer and Szabo\cite{Hummer2003}, where $P(t)$ is denoted by $S(t)$.
The model by Friddle et al.\cite{friddle_interpreting_2012} also considers small 
initial barriers, where bond reforming 
gets important and $k_r$
can not be neglected.

Further simplifications can be achieved by 
considering a constant loading rate $\alpha$\cite{garg_escape-field_1995,Hanke2006} such that the external force at a given time $t$ satisfies $F=\alpha t$. 
This assumption differs from that of a constant velocity of the 
surrounding of the bond sometimes used\cite{Dudko2003}, but should be similar if the connected cantilever and polymer spring constant $k_c$ is much smaller than the spring constant of the bond. 
This is practically always the case for the strong chemical bonds considered here as 
the spring constant of the environment $k_c$ typically involves
thousands of co-monomers including the atomic force microscopy cantilever.
This results in a very small
total effective spring constant $k_c$ , for which
the loading rate is proportional to velocity $v$ 
as $\alpha = k_\mathrm{total} v$, where $k_\mathrm{total}\approx k_c$ (for $k_c\ll k$) is spring constant of system. 
The constant loading rate then defines the time required 
to reach the maximal force  
as $t_{\mathrm{max}}=F_{\mathrm{max}}/\alpha$.

With the force given by $F=\alpha t$ at all times and setting $k_r=0$, 
we may replace $t=f t_{\mathrm{max}}$ by $f=F/F_\mathrm{max}$ and write Eq.~(\ref{eq:dPdt}) as\cite{Hanke2006}
\begin{equation}
    \frac{\partial P}{\partial f} = - t_\mathrm{max} k(f) P(f)
    = - \frac{t_\mathrm{max}}{\tau} \kappa(f) P(f) \; .
    \label{eq:Pf}
\end{equation}
Here we have defined the relative rate
\begin{equation}
   \kappa(f) = \frac{k(f)}{k(0)} = \exp\left(\beta \left[\Delta U^\ddagger - \Delta H^\ddag(f)\right]\right)
   \label{eq:kappaf}
\end{equation}
with inverse thermal energy $\beta^{-1}=k_B T$, $k_B$ denoting the Boltzmann constant and $T$ the absolute temperature.
We have expressed the rate without 
force by the theoretical lifetime of the bond $\tau=1/k(f=0)$, despite
 that this is a theoretical time only, that disregards bond-reforming. Usually $\tau\gg t_{\mathrm{max}}$.
\footnote{This assumption breaks down for very soft bonds as 
occurring in biological systems\cite{rico_heterogeneous_2019}}

Eq.~(\ref{eq:Pf}) can be formally integrated to lead to
\begin{equation}
   P(f) = P(f_0) \exp\left(-\frac{t_{\mathrm{max}}}{\tau} 
   \int_{f_0}^f \kappa(f') df' \right)
   \; ,
\end{equation}
an integral that can be analytically
solved for some special $\Delta H^\ddag(f)$. 
First to third order derivatives of $P(f)$ w.r.to $f$
can be conveniently be expressed in terms of
$P$ and $\kappa$ as given by Eqs.~(S3, S5, S7) in SI.

The  distribution $\partial P/\partial f$ determines the most probable  relative force $f^*$ measured in the experiment via\cite{williams_analytical_2003,Kempe2018,Brugner2018}
\begin{equation}
   f^* = \int\limits_0^{\infty} f' \frac{\partial P(f')}{\partial f'} df'
   \; .
   \label{eq:numerical_mpf}
\end{equation}
This function is usually highly peaked \cite{kurkijarvi_intrinsic_1972,Hanke2006}, such that
within the peaking approximation the most probable force $f^*$ will be that at the
maximum, i.e. $f$ where 
\begin{equation}
    \partial^2 P(f)/\partial f^2=0 
   \label{eq:dpdf_peak}
\end{equation}
(note that $\int_0^\infty (\partial P/\partial f) df=1$).
Eq.~(\ref{eq:dpdf_peak}) can be analytically solved for the Bell barrier from Eq.~(\ref{eq:Bell}), 
and we get\cite{evans_dynamic_1997,khodayeki_force_2022}
\begin{equation}
    f^* = \frac{1}{2 \beta \Delta U^\ddag } \text{log} \left(2 \beta \Delta U^\ddag\frac{ \tau  }{ t_\mathrm{max}} \right) \; .
    \label{eq:bell_mpf}
\end{equation}
Analytic integration is also possible the quadratic form of the barrier from Eq.~(\ref{eq:quadratic}) as detailed in Eqs.~(S10-S18) resulting in
\begin{equation}
    f^*= 1- \sqrt{\frac{1}{2 \beta \Delta U^\ddagger} W \left(
    \frac{1}{2 \beta \Delta U^\ddagger}\frac{t_{\mathrm{max}}^2}{h^2 \beta^2} \right)}\;.
    \label{eq:generalbarrier_mpf}
\end{equation}
Here $W(x)$ denotes the Lambert $W$ function.
The arguments entering Eq.~(\ref{eq:generalbarrier_mpf}) are the ratio
of the bond energy by thermal energy (through $2\beta \Delta U^\ddagger$) and
the time needed to reach $F_\mathrm{max}$ relative to the thermal attempt time to break the
bond through $t_{\mathrm{max}}/(h\beta)$.

\begin{figure}[h]
    \centering
    \includegraphics[width=\textwidth]{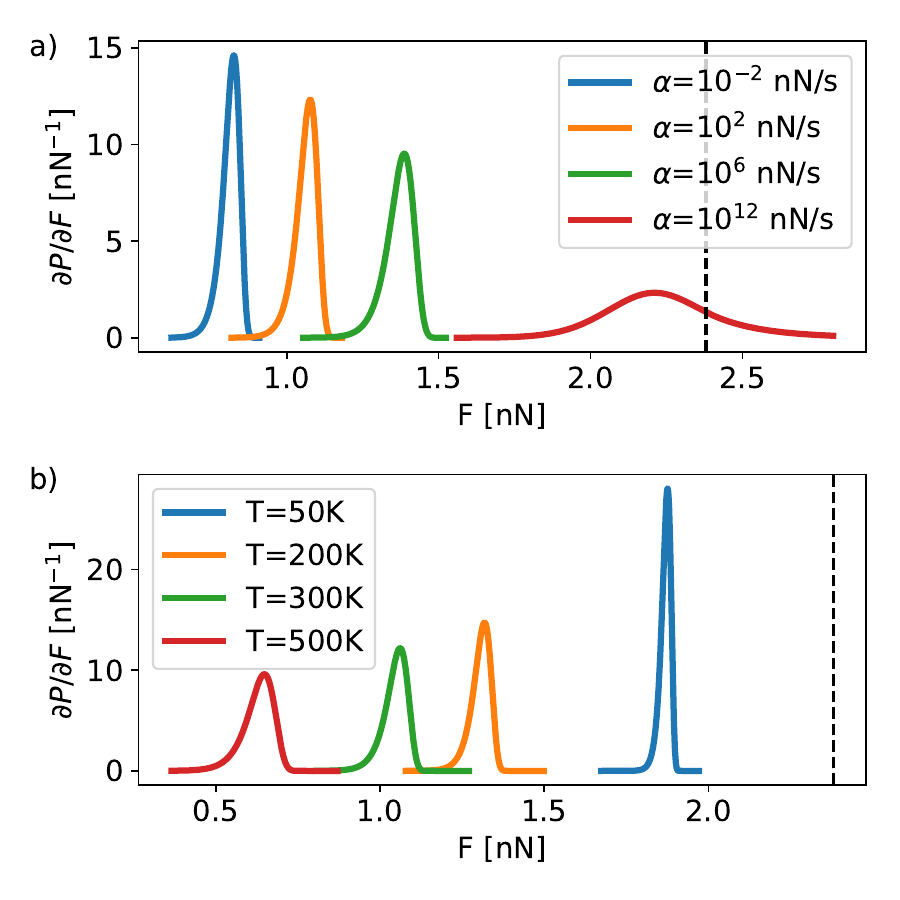}
    \caption{$\partial P/\partial F$ of the AuAg$_{2}$ molecule for a) various loading rates at 
    $T$=300 K and b) different temperatures at $\alpha$=100 nN/s.
    The broken line indicates  $F_{\mathrm{max}}$.}
\label{fig:dPdF_alpha_T}
\end{figure}
In order to get a feeling about the validity of the peaking approximation,
the distributions 
$\partial P/\partial F=F_{\mathrm{max}}^{-1 }\partial P/\partial f$ are shown  in Fig.~\ref{fig:dPdF_alpha_T}
for different loading rates 
and temperatures.
We have used the analytic
quadratic barrier Eq.~(\ref{eq:quadratic}) using $\Delta U^\ddagger$ and $F_\mathrm{max}$ 
from AuAg$_2$.
Similar distributions appear for the decay in Josephson junctions\cite{kurkijarvi_intrinsic_1972}, where an increasing external flux represents the increasing force as was noted already by Garg\cite{garg_escape-field_1995}.
These distributions are peaked in all cases and the peak position is always at 
forces smaller than $F_{\mathrm{max}}$. Increasing the loading rate (pulling speed) 
increases the mean force corresponding to the peak as there is less time 
to overcome the barrier at higher $\alpha$. 
The distribution gets significantly broader with increasing $\alpha$ as 
the potential flattens with increasing force.

The dependence on temperature is opposite to that on $\alpha$.
Increasing the temperature allows to overcome higher barriers which leads to peak positions 
at lower force. Increasing $T$ also leads to a broadening of the distribution,
such that the distributions get broader for smaller most probable forces in contrast
to the effect of increasing $\alpha$.

It could be estimated from Fig.~\ref{fig:dPdF_alpha_T} that the
peak position can also appear at negative values of $f$ if 
$\alpha$ gets very small or the temperature gets very large.
This is indeed the case if
\begin{equation}
    \frac{t_{\mathrm{max}}}{\tau} > 2 \beta \Delta U^\ddagger
\end{equation}
as seen from Eq.~(\ref{eq:bell_mpf}) directly, but can 
also be proved for Eq.~(\ref{eq:generalbarrier_mpf}) (c.f. Eq.~(S20)in SI). 
This is only the case for extremely small relative forces where the back reaction would play a significant role\cite{friddle_interpreting_2012,bullerjahn_reversible_2022}, however.

\begin{figure}[h]
    \centering
    \includegraphics[width=\textwidth]{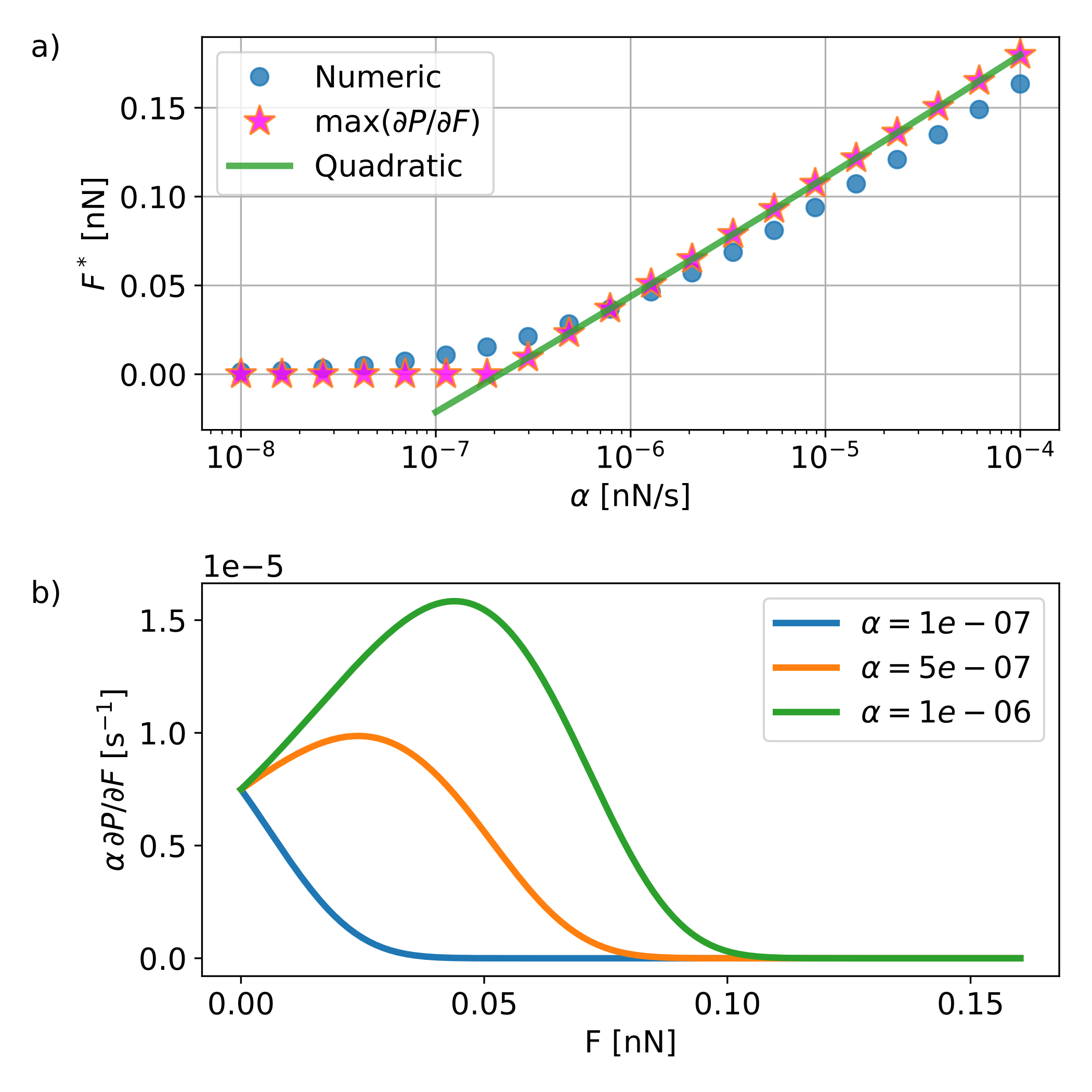}
    \caption{a) Most probable force from Eq.~(\ref{eq:numerical_mpf}) (Numeric),
    from Eq.~(\ref{eq:dpdf_peak}) (Quadratic), and from max($\partial P/\partial f$) depending on loading rate at $T$=500 K. b)
    The probability derivative distribution scaled by the loading rate $\alpha$ when the peak position
    goes towards negative values.
    }
    \label{fig:dPdF_alpha_negative}
\end{figure}
Fig.~\ref{fig:dPdF_alpha_negative} b) depicts the distribution of $\partial P/\partial F$ with decreasing loading rate $\alpha$ values. The distribution is no longer peaked in the positive force direction and thus the estimation of $f^*$
from Eq.~(\ref{eq:dpdf_peak}) breaks down as it ultimately leads to negative values. However, the most probable force from Eq.~(\ref{eq:numerical_mpf}) is strictly positive even at lower forces. 

\begin{figure}[h]  
    \centering
    \includegraphics[width=\textwidth]{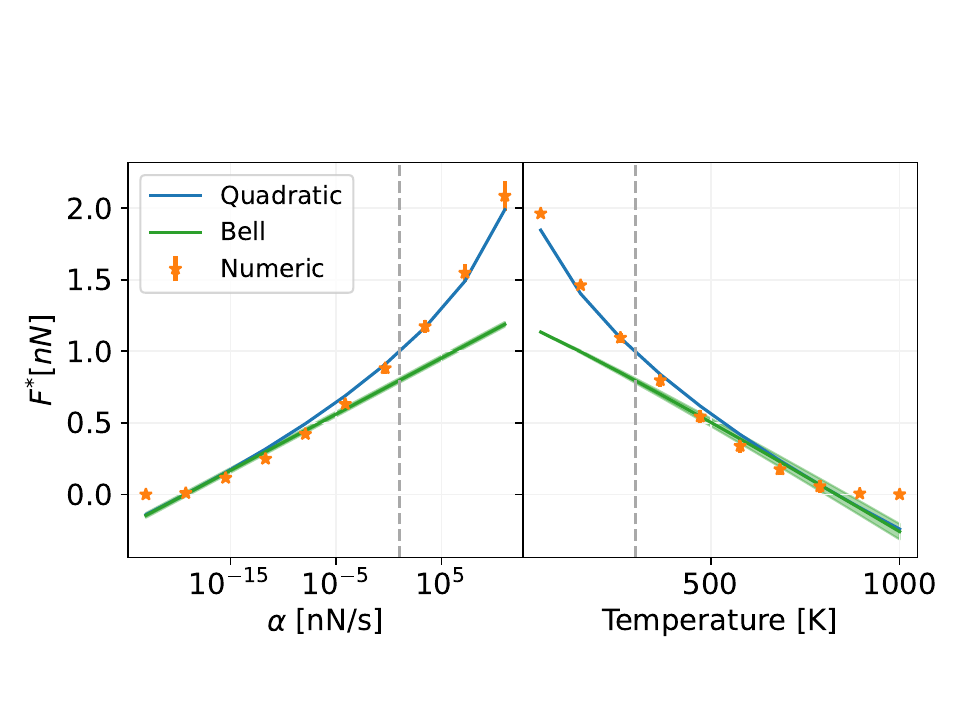}
    \caption{Most probable force for AuAg$_{2}$ molecule at various loading rates (left, for $T$ = 300 K) and temperatures (right, for $\alpha$= 10 nN/s ). 
    Eq.~(\ref{eq:numerical_mpf}) (Numeric), Eq.~(\ref{eq:bell_mpf}) (Bell) and Eq.~(\ref{eq:generalbarrier_mpf}) (Quadratic) are compared.
    }
    \label{fig:MPF_alpha_T}
\end{figure}

Fig.~\ref{fig:MPF_alpha_T} shows the most probable force for a large range of loading rates
(keeping temperature, $T$ = 300 K constant) and temperatures (keeping loading rate, $\alpha$ = 10 nN/s constant).
The errorbars for numerical values are determined
the full width at half measurement (FWHM) of the $dP/dF$ distribution. 
The shaded region indicating the error of the
the Bell prediction is obtained from Eq.~(S31). 
The analytic prediction based on the quadratic barrier closely follows the numeric integration
except of the region where negative values appear (for very low forces).
The Bell model is accurate only for low loading rates\cite{Hummer2003} and high temperatures. Hence it is useful for approximations for low forces only.
Additionally, the Bell uncertainty can only describe the increase
of the width for large temperatures, but the
dependence on the loading rate is not covered as
is explicit from Eq.~(S31). 
Most probable force and error estimates from the literature\cite{garg_escape-field_1995, dudko_extracting_2007, friddle_unified_2008, Bell1978, friddle_interpreting_2012, bullerjahn_theory_2014} are discussed in detail using a simple example in SI.

\subsection*{Comparison to single molecule force spectroscopy experiments}

We have shown above that the dissociation energy $\Delta U^\ddagger$ and the 
maximal force that the bond can withstand $F_{\mathrm{max}}$
determine the most probable force measured in the experiment through Eq.~(\ref{eq:generalbarrier_mpf}). 
The similarity of force dependent barriers with
the quadratic form seen in our toy model above and in other cases\cite{Hanke2006,khodayeki_force_2022,walter_mechanochemical_2023}
suggests that these two quantities are most important
and may be sufficient for
an accurate estimate of rupture forces measured in SMFS experiments.

\begin{figure}[h]
    \centering
    \includegraphics[width=\textwidth]{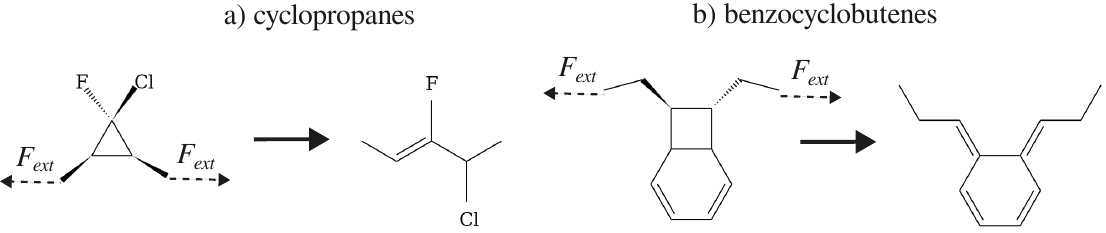}
    \caption{General reaction schemes for a) cyclopropanes and b) benzocyclobutenes.
    }
    \label{fig:reaction_shemes}
\end{figure}
We test this assumption for available experimental
data from the literature for force opening reactions of
cyclo\-pro\-panes\cite{Wang2014a, wang_reactivity_2015, klukovich_backbone_2013, wu_molecular_2010, Wang2015a} and benzocyclobutenes\cite{ Wang2015, Wang2015a}.
These types of monomers were investigated in the literature 
within SMFS experiments in many variants and 
represent non-trivial reactions involving the re-arrangements of multiple bonds. 
The force induced reactions are illustrated in Fig.~\ref{fig:reaction_shemes}, 
where the triangular cyclopropane is re-arranged to form a double and two single bonds, while the benzocyclobutenes open the
cyclobutene square to form two double bonds when forces are applied. The observed rupture forces are influenced by side
groups as well as by variation of the halogens
in cyclopropanes. 
Detailed schematics 
of all the reactions considered can be found in SI.
An extensive investigation of these compounds using
COGEF considering step size of 0.075 \AA ~did see some correlations, but was unable to get forces similar to the experimental measurements\cite{Klein2020}.

The basic ingredients we need are $\Delta U^\ddagger$ and $F_{\mathrm{max}}$, which we calculate
within density functional theory (DFT)
as implemented in the
GPAW package\cite{Mortensen05,Enkovaara10}.
GPAW applies the projector augmented wave (PAW) method\cite{Blochl_projector_1994},
where the smooth wave functions are represented on real space grids with grid spacing of
0.2 \AA, while we used grid spacing of 0.1 \AA\ for the smooth electron density.
The exchange-correlation functional is approximated as devised by Perdew, Burke and Ernzerhof (PBE)\cite{Perdew_generalized_1996}.

The question arises about the best method to evaluate $\Delta U^\ddagger$ and  $F_{\mathrm{max}}$. 
The first idea may be to use a COGEF calculation\cite{Klein2020} 
via the maximal energy and the maximal force observed.
While this approach gives good estimates for $F_{\mathrm{max}}$, 
it generally leads to an overestimation of
the dissociation energy. The overestimation is larger for a
softer elastic environment coupled to the bond, as soft elastic environments are able to absorb large amounts of energy via the applied force\cite{khodayeki_force_2022}.
This environment consists of the rest of the 
monomer-molecule surrounding the bond that breaks 
as well as co-monomers considered in
the calculation. 
A larger model (which is would appear superior at first sight) therefore worsens the effect.
In order to avoid these extra contributions,
we determine the barrier in absence of force 
$\Delta U^\ddagger=\Delta H^\ddagger(F=0)$ via nudged elastic band calculations\cite{henkelman_methods_2002,larsen_atomic_2017}.

We have considered the medium sized monomers (i.e. ring molecule and a two carbon atom chain on either side) for each molecule, as we did not observe major changes by increasing 
the monomer size further (detailed analysis is given in SI).
The stress induced pericyclic ring opening in molecules is suspected to steer the reaction in a direction away from thermally induced lower barrier reaction pathway \cite{woodward_conservation_1969}. 
These thermally activated reactions are in accordance with conservation of orbital symmetry as explained by Woodward-Hoffmann (WH)\cite{woodward_stereochemistry_1965} and
Woodward-Hoffmann-DePuy (WHD) \cite{depuy_chemistry_1968} rules. 
It was shown computationally for benzocyclobutenes\cite{Hickenboth2007, ong_first_2009} and cyclopropanes\cite{Lenhardt2010, wollenhaupt_force-induced_2018} and also 
proved experimentally \cite{Wang2014, Lenhardt2010, Wang2015} that the force induced activation may not always agree with these rules. 
In cases where forbidden reactions occur in COGEF, the expected 
product state in accordance with the WH/WHD rules was considered to calculate $\Delta U^\ddagger$. This ensures the use of the correct thermal activation barrier
(see SI for details).

Using the monomer properties $\Delta U^\ddagger$ and $F_{max}$, the missing ingredients to predict the most probable rupture force $F^*$ are the loading rate and the temperature at which the SMFS experiment is performed.
Room temperature was assumed throughout.
The experimental loading rate specified in the references is directly chosen\cite{klukovich_backbone_2013}. In cases where the values are not specified, we estimate it using the product of spring constant and experimental velocity (as $\alpha = k v$). The total spring constant of the experimental setup $k_{total}$ is obtained from a linear fit of the experimental SMFS force vs. elongation. If this calculated value results in a value larger than the given cantilever spring constant, we consider the latter for the calculation of loading rate (see Tabs.~S1 and S3 in SI for details).

\begin{figure}[h!]
    \centering
    \includegraphics[width=\textwidth]{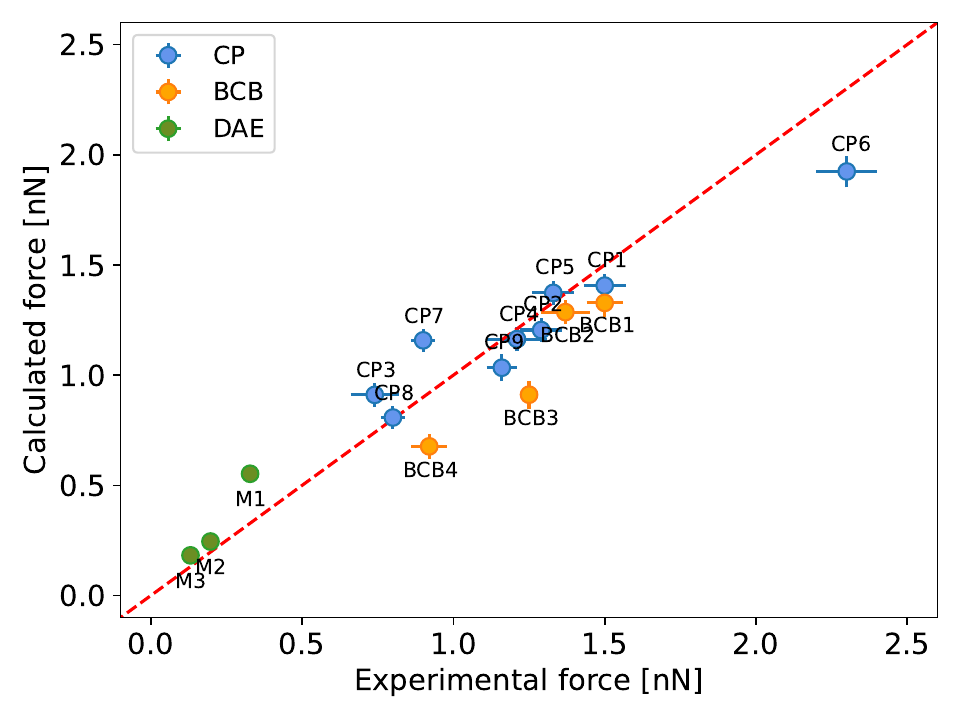}
    \caption{Comparison our our predicted most probable force to experimental results for
    cyclopropanes (CP), benzocyclobutenes (BCB) and diarylethenes (DAE). The schematics of reactions, experimental and calculated values for the molecules can be found in SI (Table S3). The meaning of the errorbars is explained in the text. The broken line indicates perfect agreement.
    }
    \label{fig:number_mpf}
\end{figure}
Using these values, we compare our predictions using Eq.~(\ref{eq:generalbarrier_mpf})
to the measured rupture forces\cite{Wang2014,klukovich_backbone_2013,faza_solvolytic_2004,Wang2015,Wang2014a,Wang2015a} in Fig.~\ref{fig:number_mpf}.
The horizontal error bars correspond to the experimental error range and the vertical error bars correspond to the prediction errors (Eq.~(S67) in SI), which is close to the width of the probability density distribution. 
We obtain very good agreement between our models and the experiment 
over a large force range. 
The prediction is quantitative and much better than the usage of 
$F_{\mathrm{max}}$\cite{Klein2020}.
Our description is thus able to predict the experimentally observed rupture forces directly from this minimal set of parameters, where only $\Delta U^\ddagger$ and 
$F_{\mathrm{max}}$ obtained from DFT enter.
 
We have also incorporated diarylethene (DAE) mechano\-phores that were investigated
in a recent study~\cite{zhang_diaryl} into the estimation 
of the most probable forces from our model. 
The DAEs were reported to undergo mechanically driven noncovalent transformations showing pronounced lever arm effects. 
We have used the COGEF $F_{\mathrm{max}}$ and the thermal barriers $\Delta U^\ddagger$
for rotation for these molecules as given in this study. 
The loading rate is estimated (0.9 nN/s for a velocity of 3000 $\text{\AA}$/s) using the cantilever spring constant from the force-extension curve similar to other molecules discussed in SI.
Inclusion of these additional molecules expands the diversity as 
comparably low forces were detected (Fig.~\ref{fig:number_mpf}).
We also get very good agreement in our prediction of the most probable force
with the measured one for this noncovalent, but still barrier-controlled
force-induced transition.

\section*{Conclusion}
\label{conclusion}

In conclusion we have derived a closed form for the prediction of
most probable bond-rupture forces as measured in  single molecule force spectroscopy experiments.
The nature of the bond exposed to external forces can be conveniently
described by two key parameters which are the bond energy in absence of force
and the maximal force that the bond can withstand.
These two quantities are usually sufficient to characterize the
force-dependent barrier until a bond breaks to good accuracy.
The approximative quadratic form assumed naturally leads to
the famous Bell model for small forces.
These necessary properties of the bond
can be obtained via density functional theory calculations
using force free transition state modeling and a COGEF calculation.

Assuming a constant force increase, the rest is statistics as the bond is
broken by thermal fluctuations. The experimentally observed
most probable breaking force is 
governed by the time scales of the force increase and 
the attempt frequency to break the bond.
This leads to a closed and simple form for the prediction
of the forces measured in experiment.

The effectiveness and accuracy of this description is tested on rupture
forces from ring opening reactions of cyclopropanes and benzocyclobutenes
as well as for diarylethene noncovalent rearrangements reported in the literature.
We find very good agreement of our predicted forces and the
experimentally observed ones in all cases.

We expect that the approximations adopted are widely applicable
also to other types of bond openings.
The approximation may break down for very complicated
barriers that can not be described by the simple quadratic form,
like retro-Diels-Alder bond openings, however\cite{walter_mechanochemical_2023}

\section*{Acknowledgements}

The authors are grateful for funding
by the Deutsche For\-schungsgemeinschaft (DFG, German Research Foundation) under
Germany’s Excellence Strategy–EXC-2193/1\-–390951807
and WA 1687/11-1.
The authors acknowledge computational resources by the state of Baden-Württemberg through
bwHPC and the German Research Foundation (DFG) through grant no
INST 39/963-1 FUGG (bwForCluster NEMO) and INST 40/575-1 FUGG (bwForCluster JUSTUS2). 

\section*{Conflict of Interest}

There are no conflicts of interest to declare.

\begin{shaded}
\noindent\textsf{\textbf{Keywords:} \keywords} 
\end{shaded}


\setlength{\bibsep}{0.0cm}
\bibliographystyle{Wiley-chemistry}
\bibliography{force_made_simple}

\clearpage


\section*{Entry for the Table of Contents}



\noindent\rule{11cm}{2pt}
\begin{minipage}{5.5cm}
\includegraphics[width=5.5cm]{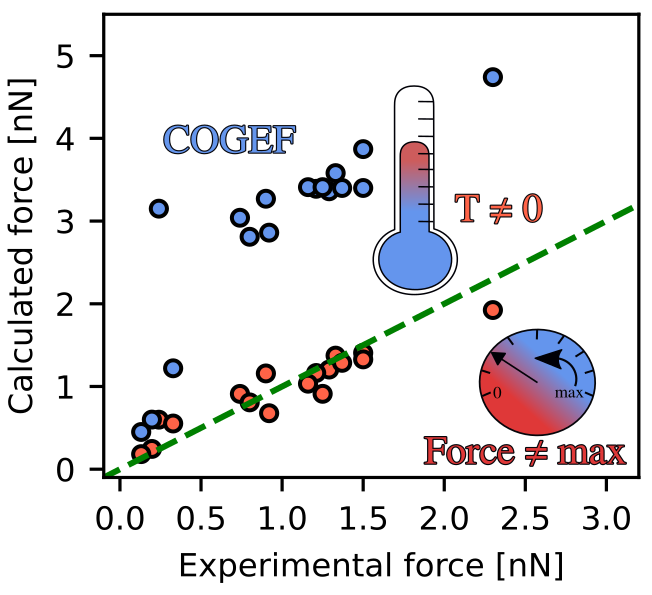} 
\end{minipage}
\begin{minipage}{5.5cm}
  \large\textsf{COGEF describes bond-breaking at zero Kelvin.
    We provide an explicit method to include finite temperature effects
    pedicting most-probable bond-opening forces as seen in experiment.
  }
\end{minipage}
\noindent\rule{11cm}{2pt}

\vspace{2cm}

\end{document}